\documentstyle[aps,prl,twocolumn,psfig,floats]{revtex}     

\begin{document}
\preprint{\vbox{\hbox{{\tt hep-ph/0207009}\\ June 2002}}}
\draft
\wideabs{
\title{Kaluza-Klein States versus Winding States:\\
        Can Both Be Above the String Scale?} 
\author{Keith R. Dienes \,and\, Arash Mafi}
\address{Department of Physics, University of Arizona, Tucson, AZ  85721 USA}
\address{E-mail addresses:  ~{\tt dienes,mafi@physics.arizona.edu}}
\date{June 28, 2002}
\maketitle
\begin{abstract}
     When closed strings propagate in extra compactified dimensions,
     a rich spectrum of Kaluza-Klein states and winding states
     emerges.  Since the masses of Kaluza-Klein states and winding states 
     play a reciprocal role,
     it is often believed that either the lightest Kaluza-Klein states
     or the lightest winding states must be at or below the string scale.
     In this paper, we demonstrate that this conclusion is no longer true for
     compactifications with non-trivial shape moduli.  Specifically, we 
     demonstrate that toroidal compactifications exist for which all 
     Kaluza-Klein states as well as all winding states are 
     heavier than the string scale.  This observation could have important
     phenomenological implications for theories with reduced string scales,
     suggesting that it is possible to cross the string scale without
     detecting any states associated with spacetime compactification.
\end{abstract}
\bigskip
\bigskip
          }

\newcommand{\newc}{\newcommand}
\newc{\gsim}{\lower.7ex\hbox{$\;\stackrel{\textstyle>}{\sim}\;$}}
\newc{\lsim}{\lower.7ex\hbox{$\;\stackrel{\textstyle<}{\sim}\;$}}

\def\beq{\begin{equation}}
\def\eeq{\end{equation}}
\def\beqn{\begin{eqnarray}}
\def\eeqn{\end{eqnarray}}
\def\calM{{\cal M}}
\def\Nbar{{\overline{N}}}
\def\half{{\textstyle{1\over 2}}}
\def\ie{{\it i.e.}\/}
\def\eg{{\it e.g.}\/}
\def\etc{{\it etc}.\/}


\def\inbar{\,\vrule height1.5ex width.4pt depth0pt}
\def\IR{\relax{\rm I\kern-.18em R}}
 \font\cmss=cmss10 \font\cmsss=cmss10 at 7pt
\def\IQ{\relax{\rm I\kern-.18em Q}}
\def\IZ{\relax\ifmmode\mathchoice
 {\hbox{\cmss Z\kern-.4em Z}}{\hbox{\cmss Z\kern-.4em Z}}
 {\lower.9pt\hbox{\cmsss Z\kern-.4em Z}}
 {\lower1.2pt\hbox{\cmsss Z\kern-.4em Z}}\else{\cmss Z\kern-.4em Z}\fi}

\input epsf


\section{Introduction}

Theories involving large extra spacetime dimensions and fundamental
theories at reduced energy scales have attracted considerable attention
in recent years~\cite{TeV1,TeV2,TeV3,TeV4}.  As a result of such developments, it 
has become possible to contemplate experimental probes not only of 
compactification geometry, but also of quantum gravity, grand unification, 
and even string theory.  Most recent phenomenological studies of such 
theories have focused on the effects of their low-energy Kaluza-Klein 
states.  However, if the string scale itself is in the TeV range, then
many additional string states may also play a role in the effective 
low-energy phenomenology.    

In general, closed string theories (as well as certain closed-string
sectors of Type~I open string theories) give rise not only to Kaluza-Klein
states, but also to string winding states as well as string oscillator states.
These states may be differentiated by the manner in which their masses
depend on the scale of compactification as well as on
the string tension (equivalently the string scale).
Kaluza-Klein states reflect the higher-dimensional
momentum quantization induced by
the spacetime compactification,
and consequently have masses inversely proportional
to the compactification radius but independent of the string scale.
By contrast, winding states reflect the possibility that the
string can stretch completely around the compactified 
dimension.  These states therefore have masses that grow
linearly with the compactification radius as well as with the
string tension.
Finally, string oscillator states correspond
to the vibrational modes of the fundamental string.
Their masses are thus set purely by the 
string tension (or equivalently 
the fundamental string scale).
Note that while the masses of the Kaluza-Klein and winding states are sensitive
to the compactification geometry, the spectrum of oscillator states
is completely independent of this geometry.
Likewise, the masses of the Kaluza-Klein excitations are independent
of the string scale, whereas the masses of the winding and oscillator 
states depend critically on the string scale.

There are, of course, many instances in which it is legitimate
to restrict our attention to various subsets of these states.  For example,
in theories whose fundamental constituents
are point particles, there are no winding or oscillator states.
The Kaluza-Klein spectrum therefore provides a complete low-energy
description of the
resonances associated with spacetime compactification.  
Likewise, even within the context of string theory, 
it is again legitimate to restrict our attention
to the Kaluza-Klein states
if the compactification radius is significantly larger than
the fundamental string length.  
Indeed, in such cases, the winding states
will be much heavier than the string scale, and will
affect phenomenology only at correspondingly higher energies. 

However, if the string scale is in the TeV range,
some or all of
the compactification radii may not differ significantly 
from the string scale~\cite{TeV1,TeV3,TeV4}.  In such cases, Kaluza-Klein states,  
winding states, and even oscillator states must be considered
together.  
Their combined effects can therefore give
rise to dramatic alternatives to 
traditional weak-scale supersymmetric or Kaluza-Klein 
approaches to the gauge hierarchy problem~\cite{missusy}.
In such cases, however, an important question
is to determine which of these states are truly
the lightest in the string spectrum.  In other words,
which states can be expected to appear below, at, and above the string scale?

Ordinarily, the masses of Kaluza-Klein states and winding states play a 
reciprocal role:  if the lightest Kaluza-Klein states are lighter than the string
scale, then the corresponding winding states are necessarily
heavier than the string scale.  Similarly, the reverse situation in which 
the lightest Kaluza-Klein states are heavier than the string scale
ordinarily results in winding states which are lighter than the
string scale (and is equivalent to the previous configuration 
as a result of $T$-duality).
The expectation, then, is that {\it either}\/ Kaluza-Klein 
states {\it or}\/ the winding states must 
be lighter than the string scale, or must at least have masses
equal to the string scale.  Thus, it would seem that 
it is impossible to cross the string scale without seeing at
least some states (either Kaluza-Klein or winding) associated
with the compactification geometry.
 
In this paper, we shall demonstrate that
this na\"\i ve expectation is incorrect in the case
of compactifications with non-trivial shape moduli.  
Specifically, we
shall show that it is possible for the string scale to be 
simultaneously {\it lighter}\/ than {\it all}\/ the Kaluza-Klein states 
as well as {\it all}\/ the winding states.
Thus, in such theories, it is possible to cross the string scale
without seeing a single resonance associated with the spacetime 
compactification!  
Needless to say, this can therefore give rise to a 
a low-energy phenomenology which 
is markedly different from that of the usual 
Kaluza-Klein effective field theories.

We shall begin by focusing
purely on the spectrum of Kaluza-Klein and winding modes
associated with toroidal compactification.  After establishing
our main result, we shall then proceed to discuss how this
result emerges within the context of a full string theory 
when the string oscillator states are also taken into account.

\section{Spectrum of Compactification Resonances}

We begin by considering the action for a string propagating
in an $n$-dimensional spacetime governed by a metric $G_{ij}$
and a background antisymmetric tensor (torsion) field $B_{ij}$:  
\beqn
   {\cal S} &=& - {1\over 4\pi\alpha'} \int d^2\sigma 
    \, \biggl(\sqrt{-g}g^{\alpha\beta}G_{ij}\partial_\alpha X^i
    \partial_\beta X^j \nonumber\\
     && ~~~~~~~+~\epsilon^{\alpha\beta}B_{ij}\partial_\alpha X^i
    \partial_\beta X^j\biggr)~.
\label{action}
\eeqn
Here $\alpha'\equiv M_{\rm string}^{-2}$ is the string Regge slope,
and $\sigma^\alpha$,  $\partial_\alpha$, and $g_{\alpha \beta}$ are the 
coordinates, derivatives, and metric on the string worldsheet.
The spacetime indices $i,j$ run from $1,...,n$.
Because our interest is in the 
spacetime compactification of this theory, we have restricted the 
above action
to include only those bosonic worldsheet fields $X^i$
corresponding to the spacetime coordinates to be compactified.  Later we 
will discuss how this may be embedded in a more complete string
theory.

The next step is to compactify these $n$ spacetime dimensions.
For simplicity, we shall take our compactification manifold
to be a general, flat, $n$-dimensional
torus.  Such a torus can be specified by its periodicity radii 
$R_i$ ($i=1,...,n$) as well as by
shape angles $\alpha_{ij}$ ($i,j=1,...,n$).  These
shape angles 
parametrize the relative orientations 
between the $i^{\rm th}$ and $j^{\rm th}$ toroidal periodicities.

Quantizing the string in the usual fashion,
we then find that 
the resulting string states can be classified 
in terms of their Kaluza-Klein  momentum numbers $n_i$ and
their winding numbers $w_i$, where $i=1,...,n$.
If we collect these momentum and winding numbers into a vector
\beq
     \tilde{N}^T~ =~\left(\frac{n_1}{R_1},...,\frac{n_n}{R_n};
     \frac{w_1R_1}{\alpha^\prime},...,\frac{w_nR_n}{\alpha^\prime}\right)~,
\label{KK}
\eeq
we find that the mass of the corresponding state takes the 
form~\cite{Giveon,nextpaper}
\beq
       {\cal M}^2_{n_i;w_i}~=~\tilde{N}^T Q^{-1} \tilde{N}
\label{masswithB}
\eeq
where
\beq
       Q^{-1}~ =~  \pmatrix{  G^{-1} & G^{-1}B \cr
                                -B G^{-1} & G - BG^{-1} B \cr}~.
\label{qminusone}
\eeq
In this expression, $G$ is the dimensionless $(n\times n)$-dimensional 
metric of the $n$-torus,
now given by 
\beq
                  G_{ij}~\equiv~ \cos\alpha_{ij}~
\label{Gdef}
\eeq
(where $\alpha_{ii}\equiv 0$),
and $B$ is the $(n\times n)$-dimensional background antisymmetric tensor
as in Eq.~(\ref{action}).
The result in Eq.~(\ref{masswithB}) is completely standard~\cite{Giveon};  
a full derivation will be given in Ref.~\cite{nextpaper}.

Let us now examine specific examples of this mass formula.
For a one-dimensional torus (\ie, a circle), there are no shape angles $\alpha_{ij}$.
There is also no $B$-field background.  We then obtain the standard
spectrum associated with circle-compactification:
\beq
        {\calM}_{n_1;w_1}^2 ~=~  {n_1^2\over R^2} + {w_1^2 R^2 \over \alpha'^2}~.
\label{circlestates}
\eeq
Note that this spectrum conforms to our usual expectations:  depending on
the value of the radius $R$, either the 
momentum (Kaluza-Klein) or the winding modes must be lighter than the string 
scale $M_{\rm string}\equiv 1/\sqrt{\alpha'}$.
Specifically, the mass of the lightest Kaluza-Klein and winding states
are respectively given by
\beq
      \calM^2_{\rm KK}= {1\over R^2}~,~~~~~
      \calM^2_{\rm winding}= {R^2 \over \alpha'}~.
\eeq
The self-dual radius $R_\ast$ (defined to be the radius at which these 
masses become equal)
is therefore given by $R_\ast \equiv \sqrt{\alpha'}$,
implying that
\beq
       M_{\rm string} ~=~ {1\over R_\ast} ~=~ M_\ast~
\label{circleresult}
\eeq
where $M_\ast\equiv \calM_{\rm KK} = \calM_{\rm winding}$
at the self-dual radius.
These results are completely as expected, and it is
possible to exploit $T$-duality in order to choose a convention
such that $R^{-1} < M_{\rm string}$ (or equivalently 
$\calM_{\rm KK} < \calM_{\rm winding}$)
when we are not at the self-dual radius.
  
However, as we shall now demonstrate, these results no longer hold
in the case of higher-dimensional toroidal compactifications with non-trivial
shape moduli and background $B$-fields.
In the case of a two-torus, the antisymmetric $B$-field has a single element 
$b\equiv B_{12}$;  there is also only one shape modulus $\theta\equiv \alpha_{12}$.  
The corresponding mass formula then takes the form
\beqn
     \calM_{\vec n; \vec w}^2 &=& {1\over \sin^2\theta}
     \left( {n_1^2\over R_1^2} + {n_2^2\over R_2^2}
           - 2 {n_1n_2\over R_1R_2} \cos\theta\right)\nonumber\\
     && + {{b^2+\sin^2\theta}\over {\alpha^{\prime 2}\sin^2\theta}}
     \left( {w_1^2 R_1^2} + {w_2^2 R_2^2} + 2 {w_1w_2 R_1R_2} \cos\theta\right)\nonumber\\
     && + {2b\over \alpha' \sin^2\theta}
            \biggl( n_1 w_1\cos\theta - n_2 w_2 \cos\theta \nonumber\\
     && ~~~~~~~~~~~~~~~ + n_1w_2{R_2\over R_1} - n_2 w_1{R_1\over R_2} \biggr)~.
\label{Kaluza-Klein}
\eeqn
Without loss of generality, let us take $R_1\geq R_2$.
The lightest Kaluza-Klein state then corresponds to the state
with $(n_1,n_2;w_1,w_2)= (1,0;0,0)$, and has mass
\beq
        \calM_{\rm KK} ~=~ {1\over R_1 \sin\theta}~.
\eeq
By contrast, the lightest winding mode 
in the same direction associated with $R_1$ 
is the state $(n_1,n_2;w_1,w_2)= (0,0;1,0)$,
with mass
\beq
        \calM_{\rm winding} ~=~ 
          {\sqrt{b^2+\sin^2\theta}\over \alpha' \sin\theta} \,R_1~.
\eeq
Solving for the self-dual radius, we find
\beq
       R^2_{\ast} ~=~ {\alpha'\over \sqrt{b^2 + \sin^2\theta}}~.
\eeq
Thus, defining as before $M_\ast\equiv \calM_{\rm KK} = \calM_{\rm winding}$
at the self-dual radius,
we obtain the result 
\beq
       M^2_{\rm string} ~=~ {1\over \sqrt{b^2+\sin^2\theta}} \, {1\over R^2_{\ast}} 
              ~=~ {\sin^2\theta \over \sqrt{b^2+\sin^2\theta}} \, M_\ast^2~.
\label{twotorusresult}
\eeq
This is to be compared with the case for a one-torus in Eq.~(\ref{circleresult}).

Clearly, if $\theta=\pi/2$ and $b=0$, we obtain the expected result
that $M_\ast = M_{\rm string}$.
In this case, it is impossible to make the Kaluza-Klein states heavier
than the string scale without making at least one winding state
lighter than the string scale.
However, in all other cases, we find from Eq.~(\ref{twotorusresult})
that 
\beq
         M_\ast ~>~ M_{\rm string}~!
\eeq
This implies that it is possible (\eg, at the self-dual radius) for
the string scale to be lighter than {\it all}\/ the Kaluza-Klein modes
as well as {\it all}\/ the winding modes!  
Indeed, it is possible to cross the energy threshold associated
with the string scale without having seen a single compactification 
resonance of either type!
Choosing $R_2=R_{\ast}$ in this situation
then guarantees that {\it all}\/ compactification resonances associated
with this two-torus compactification will be heavier than the string scale.
 
This feature generalizes to higher-dimensional compactifications.
In the case of a three-torus, for example, there are three radii $R_i$,
three shape angles $\lbrace \alpha_{12}, \alpha_{13}, \alpha_{23}\rbrace$,
and likewise three independent components $\lbrace b_{12}, b_{13}, b_{23}\rbrace$
for the background antisymmetric $B$-field.
Note that the shape angles must satisfy the constraint
\beq
       |\alpha_{12} - \alpha_{13}|  ~<~ \alpha_{23} ~<~ \alpha_{12} + \alpha_{13}~
\label{constraint}
\eeq
in order to guarantee that our three-torus is physical;  
these inequalities are saturated only in the degenerate limit when the direction associated
with the third toroidal periodicity lies in the plane formed by the other two. 
Repeating the above procedure to calculate the Kaluza-Klein and winding masses
associated with the first toroidal periodicity, 
we find that Eq.~(\ref{twotorusresult})
becomes
\beq
     M_{\rm string}^2 ~=~ {s_{23}\over \sqrt{K+K_1}} {1\over R_{1\ast}^2} ~=~
                 {K\over s_{23} \sqrt{K+K_1}}\, M_{1\ast}^2~
\label{threetorusresult}
\eeq
where $s_{ij}\equiv \sin\alpha_{ij}$, $c_{ij}\equiv\cos\alpha_{ij}$,
and where
\beqn
        K&\equiv& \det\, G ~=~ 1-c_{12}^2 - c_{13}^2 - c_{23}^2 + 
                       2 c_{12} c_{13} c_{23}~,\nonumber\\
        K_1 &\equiv&  b^2_{12} s_{13}^2 + b_{13}^2 s_{12}^2 - 
                      2 b_{12} b_{13} (c_{23} - c_{12} c_{13})~.
\label{Kdefs}
\eeqn
Similar results hold for the remaining toroidal periodicities.
Note that this result reduces to Eq.~(\ref{twotorusresult}) in the special
case with $c_{13}=c_{23}=b_{13}=b_{23}=0$.
 
Several features of this result are immediately apparent.
First, it is straightforward to verify that 
\beq
             {K\over s_{23} \, \sqrt{K+K_1}} ~ \leq~1 
\eeq
whenever the shape angles satisfy the constraint in Eq.~(\ref{constraint}). 
Indeed, we find that $K$ and $K_1$ are both necessarily positive when
Eq.~(\ref{constraint}) is satisfied.
Thus, once again, we verify that 
the self-dual Kaluza-Klein/winding mass scale is 
greater than $M_{\rm string}$.
Indeed, this holds for all three periodicities of the torus.
 
More interestingly, however, we now observe a new feature:  the self-dual radius,
and indeed the mass of the lightest Kaluza-Klein/winding modes at the self-dual
radius, are not universal for all toroidal directions!
Instead, they depend on the specific configuration of shape angles and
antisymmetric $B$-field components involved in the compactification.
Thus, the whole notion of self-dual radius becomes a shape-dependent
phenomenon, varying according to the specific direction of compactification.

\section{Geometric Bounds}

Thus far, we have demonstrated that in compactifications
with non-trivial shape moduli, $M_\ast$ can generically be larger
than $M_{\rm string}$.
As we have seen, this implies that compactifications exist for
which the lightest Kaluza-Klein states as well as the 
lightest winding states are simultaneously heavier than the string scale.

Given this result, it is natural to wonder how large this separation
between the lightest compactification states and the string scale can become.
In other words, what is the maximum size for the ratio $M_\ast/M_{\rm string}$?

At first glance, it may appear that this ratio is completely unbounded.
For example, in the case of a two-torus in Eq.~(\ref{twotorusresult}),
it might initially appear that we can take $\theta\ll 1$, thereby
making $M_\ast$ arbitrarily heavy.  However, there is an important
subtlety that must be taken into account.  In our derivation of
the two-torus result in Eq.~(\ref{twotorusresult}), we assumed that the lightest
Kaluza-Klein and winding states are those with $(n_1,n_2;w_1,w_2)=(1,0;0,0)$
and $(0,0;1,0)$ respectively.
However, this assumption is true only if two conditions are satisfied.
First, this is true only if $|\cos\theta|$ remains relatively small, so that
no anomalous cancellations are induced in the mass formula in Eq.~(\ref{Kaluza-Klein})
when $n_1,n_2$ are both non-zero and large.
Or, to phrase this restriction more mathematically in the case
of a two-torus, we must ensure
that $\tau\equiv (R_2/R_1)e^{i\theta}$ remains within the fundamental
domain of the modular group.
Second, 
just as the radii and shape angles must be restricted by modular symmetries,
we must also restrict the components of the antisymmetric $B$-field.
It turns out~\cite{Giveon} that the mass spectrum has a symmetry under which
the components of $\hat B_{ij}\equiv (R_i R_j) B_{ij} /\alpha'$
are each individually shifted by integers.
This implies that we must restrict each of the $B$-field components
such that $-1/2 < \hat B_{ij} \leq 1/2$ for all $i,j$.

Let us examine these constraints for the two-torus.
In this case, it is easy to verify that 
the maximum ratio $M_\ast/M_{\rm string}$ is
achieved for the so-called $SU(3)$ torus: 
\beq
        R_1=R_2=R_\ast=\sqrt{\alpha'}~,~~~~~ \cos\theta=b=1/2~,
\label{SU3torus}
\eeq
yielding $M^2_\ast/M^2_{\rm string}= 4/3$.
Note that this solution 
satisfies $R_\ast=\sqrt{\alpha'}$.   This implies
that $\hat B=B$, thereby guaranteeing that 
this solution is also consistent with the above $B$-field constraint.
 
Even greater ratios can be achieved for higher-dimensional
compactifications.  For example, if we perform a five-dimensional
compactification on the so-called $SO(10)$ torus given by
\beq
    \cases{  R_i = \sqrt{\alpha'} ~~{\rm for~all}~i~,\cr
      \alpha_{12}=\alpha_{23}=\alpha_{34}=\alpha_{35} = \pi/3~, \cr 
        b_{12}=b_{23}=b_{34}=b_{35} = 1/2~ \cr}
\label{SO10torus}
\eeq
(with all other $\alpha_{ij}=\pi/2$ and $b_{ij}=0$), we obtain
the ratio
$M^2_\ast/M^2_{\rm string}= 2$ for each of the five compactified
dimensions.
Thus, in this case, the lightest Kaluza-Klein and winding states
do not appear until at least the {\it second}\/ excited level.
Of course, these particular tori are chosen merely as highly
symmetric illustrative examples, and it is 
likely that more dramatic ratios can be achieved in asymmetric 
compactifications for which each compactified
dimension has its own value of $M_\ast$.  We have already seen this possibility,
for example, in the three-torus case discussed above.
These issues will be discussed further in Ref.~\cite{nextpaper}.

\section{Embeddings into String Theory}

We now discuss the additional features that arise when 
these results are embedded into the full string framework.
As we shall see, this is important because 
there are significant differences between
the traditional Kaluza-Klein picture and its 
ultimate realization in string theory.

The full mass spectrum of a toroidally compactified string
is generally governed by two equations of the form~\cite{reviews}
\beqn
  \alpha' \calM_{\rm tot}^2  &=&  \alpha' \calM_{n_i;w_i}^2  
        +  2\,(N+\Nbar) + 2(a + \overline{a}) \nonumber\\
  \Nbar-N &=&  \sum_i n_i w_i + a - \overline{a} ~.
\label{constraints}
\eeqn
Here $\calM_{n_i;w_i}^2$ represents
the mass contribution from the Kaluza-Klein and winding
excitations, as defined in Eq.~(\ref{masswithB}),
and $(N,\Nbar)$ are the total energies from the left- and right-moving
string oscillator excitations.  These energies are calculated as
$N\equiv \sum_{n} n N_n$
and $\Nbar\equiv \sum_{n} n \Nbar_n$,
where $N_n$ and $\Nbar_n$ count the number of excitations
of frequency $n$ of the underlying left- and right-moving worldsheet fields.
Likewise, $(a,\overline{a})$ represent left- and right-moving 
vacuum-energy contributions
which we shall keep arbitrary for now.
The first of these equations indicates that the total mass of a given
string state receives contributions not only from 
Kaluza-Klein and winding excitations, but also from the oscillators
and from the total vacuum energy.
By contrast, the second of these equations is a {\it level-matching}\/
constraint which is required for the self-consistency of the string.
This constraint ensures that the total mass of a given state 
receives equal contributions
from the left- and right-moving worldsheet excitations, possibly offset
by a vacuum energy difference $\overline{a}-a$.
(The quantity $\sum_i n_i w_i$ represents the energy offset 
due to the Kaluza-Klein/winding modes.)
While there may be additional GSO constraint equations governing the string
spectrum in the case of realistic string models~\cite{reviews,phenoreviews}, 
the constraints listed in Eq.~(\ref{constraints}) are generic and appear
as a minimal set for all toroidally compactified string theories.

Given the constraints in Eq.~(\ref{constraints}),
we immediately see two important differences relative to the traditional
Kaluza-Klein picture which focuses only on $\calM_{n_i;w_i}$.  First,   
we see that the total spacetime masses of our Kaluza-Klein and winding modes
are generally offset by a non-zero vacuum energy $a+\overline{a}$.  The value 
of $a+\overline{a}$
depends on the type of string in question (bosonic, superstring, \etc), 
as well as on the type of sector (Neveu-Schwarz, Ramond, higher-order twists, \etc)
within a given string theory.  The important point, however, is that
$a+\overline{a}$ can have either sign, and is often negative in many string sectors.  
This implies that 
the lightest Kaluza-Klein states according to the mass formula
in Eq.~(\ref{masswithB}) need not be the 
lightest Kaluza-Klein states in the actual string spectrum.
We shall see several dramatic examples of this below.

The second important feature is the presence of a level-matching
constraint which {\it correlates}\/ the Kaluza-Klein/winding numbers $(n_i;w_i)$
with the string oscillator energies $(N,\Nbar)$.
This implies that whether a given Kaluza-Klein or winding state
is allowed in string theory 
depends not only on its total mass, but also on the particular
combination $\sum_i n_i w_i$.  

In order to 
see how these two features affect the mass spectrum, 
let us classify the various string states into
sectors according to their values of $N+\Nbar + a+\overline{a}$.

In sectors with $N+\Nbar + a+\overline{a}>0$, the string states are
already massive even before any Kaluza-Klein or winding modes are excited.
Such states are therefore not likely to be among the lightest states
in the full string spectrum.

Next, we turn our attention to string sectors with $N+\Nbar + a+\overline{a}=0$.
Since this implies that $\calM_{\rm tot} = \calM_{n_i;w_i}$,
our previous results apply directly in such sectors.
In other words, we have shown that there exist toroidal
compactifications such that 
the Kaluza-Klein and winding states in such sectors are all heavier
than the string scale.
Note that the special case with $n_i=w_i=0$ (corresponding
to the absence of any Kaluza-Klein or winding excitations) 
corresponds to the massless string states (such as the graviton) which already 
exist in the spectrum prior to compactification.
The remaining cases with non-zero $n_i$ and $w_i$ thus correspond to the
usual Kaluza-Klein and winding excitations of these states,
as well as the excitations of their Kaluza-Klein descendants
[including the $U(1)$ graviphoton gauge fields which emerge
from the metric and antisymmetric tensor via the original Kaluza-Klein mechanism].
It is important to note, however, that the spectrum of Kaluza-Klein and
winding excitations in such sectors is restricted 
to those states with a common value of $\sum_i n_i w_i$ (typically
zero since the lightest Kaluza-Klein states in such sectors
have $\sum_i n_i w_i=0$). 

Finally, we must consider the sectors with $N+\Nbar+a+\overline{a}<0$.
In such sectors, the lightest Kaluza-Klein and winding excitations
are tachyonic and must be GSO-projected out of the spectrum.
Depending on the specific string model in question,
there are two possible results of such GSO projections.
First, it may happen that this entire sector 
of the string theory is GSO-projected out of the spectrum.
In such cases, no further considerations are necessary.
On the other hand, it may happen that these GSO projections
affect only the lightest states, preserving
states with $\alpha' \calM_{\rm tot}^2 \geq 0$.
In such cases, then, we find that it is the {\it multiply excited}\/ 
Kaluza-Klein/winding states in these sectors which are actually the lightest states
which appear in the string spectrum!

As an explicit example of this phenomenon,
let us consider the bosonic string
(for which $a=\overline{a}= -1$).
If $(N,\Nbar)=(1,0)$ or $(0,1)$,
then $\alpha' \calM_{\rm tot}^2 = \alpha' \calM_{n_i;w_i}^2 -2$.
Thus, Kaluza-Klein or winding states for which $\alpha' \calM_{n_i;w_i}^2 =2$
are actually massless.   
For example, if two dimensions in this theory are compactified on the 
so-called $SU(3)$ torus in Eq.~(\ref{SU3torus}), then
the following twelve multiply excited Kaluza-Klein/winding
states $(n_1,n_2;w_1,w_2)$ have $\alpha'M_{n_i;w_i}^2=2$ and hence are 
massless:
\beqn
   &&  \pm (1,1;1,0),~\pm (1,0;1,-1),~\pm (0,1;0,1)\nonumber\\
   &&  \pm (1,1;0,-1),~\pm (1,0;-1,0),~\pm (0,1;1,-1)~.
\label{SU3gaugebosons}
\eeqn 
Note that the first six states [top line of Eq.~(\ref{SU3gaugebosons})] have
$\sum_i n_i w_i= 1$, while the second six states 
[bottom line of Eq.~(\ref{SU3gaugebosons})]
have  $\sum_i n_i w_i= -1$.
Thus the first group of states can have $(N,\Nbar)=(0,1)$, while the
second group can have $(N,\Nbar)=(1,0)$.
Moreover,
if these oscillator excitations correspond to excitations of the 
spacetime coordinates of uncompactified dimensions,
then these states are spacetime vectors.

Clearly, massless vectors must correspond to gauge bosons.
Indeed, it turns out that these twelve states combine with the four $U(1)$ graviphoton  
states which emerge from the Kaluza-Klein decomposition of the graviton
and $B_{\mu \nu}$ field, thereby enhancing    
the total Kaluza-Klein gauge symmetry 
from $U(1)^4$ to $SU(3)_L \times SU(3)_R$!
[It is for this reason that the torus in Eq.~(\ref{SU3torus}) is called an $SU(3)$ torus.]
Likewise, a five-dimensional compactification on
the five-dimensional torus of Eq.~(\ref{SO10torus}) produces
eighty multiply excited states with $\alpha' M_{n_i;w_i}^2=2$ and $\sum_i n_i w_i=\pm 1$,
thereby enhancing the total Kaluza-Klein gauge symmetry from $U(1)^{10}$  
to $SO(10)_L\times SO(10)_R$.

Of course, this is nothing but the standard Narain mechanism by which 
one obtains non-abelian, simply laced, level-one affine 
gauge symmetries via toroidal compactifications
in string theory:  the Kaluza-Klein/winding quantum numbers  become
identified as the charges of a non-abelian gauge group~\cite{reviews,Narain}.
(The analogous production of non-simply laced and higher-level affine gauge groups  
is discussed in Ref.~\cite{higherlevel}.)
Likewise, similar results hold for the superstring and the heterotic string. 

These features indicate that
the simple Kaluza-Klein effective field-theory picture becomes far richer,
but also far more complex, when embedded within the full context of string theory.
For example, if the $SU(3)$ gauge symmetry of the strong interaction
is realized in string theory as an enhanced gauge symmetry, as described above, 
then the massless gluons of the Standard Model are actually simultaneous
combinations  of Kaluza-Klein, winding, and oscillator modes!
In other words, such gluons correspond to string states which
simultaneously resonate in extra dimensions (Kaluza-Klein), 
wrap around those extra dimensions
(winding), and also vibrate along their length with a certain frequency (oscillators).
Likewise, the masses of the $W^\pm$ and $Z$ gauge bosons, ordinarily 
generated through the Higgs mechanism,  can equivalently be generated
by shifting various compactification radii away from the symmetric
values which are needed to produce unbroken $SU(2)\times U(1)$ gauge symmetry.

Given these observations, we see that we must be very careful when interpreting
our results within the context of string theory.
As an example, let us again consider the case of the bosonic  
string with two dimensions compactified on the $SU(3)$ torus.  
We have already shown that such a toroidal compactification
has $M_\ast^2/M_{\rm string}^2 = 4/3$.
However, the presence of negative vacuum energies in the bosonic
string requires that we must consider the higher Kaluza-Klein/winding
excitations as well.
It turns out that the complete spectrum
of Kaluza-Klein/winding modes on this torus consists of
\beq
     \alpha' M_{n_i;w_i}^2 ~=~ 0,~4/3,~2,~ 10/3,~4,~16/3,~...
\label{SU3spectrum}
\eeq
with values
\beq
      \sum_i n_i w_i  ~=~ 0,~ 0,~\pm 1,~\pm 1,~0,~\lbrace 0,\pm 2\rbrace,~...
\label{SU3spectrumndw}
\eeq
respectively.
[The states at $\alpha' M_{n_i;w_i}^2=2$ are the gauge boson
states listed in Eq.~(\ref{SU3gaugebosons}).]
Likewise, there are only six sectors of the bosonic string
which can possibly contain light states (defined as states
with $\alpha' M_{\rm tot}^2 <2$):  these are 
the sectors with $(N,\Nbar)= (0,0), (1,1), (1,0), (0,1), (2,0)$, 
and $(0,2)$.
The allowed Kaluza-Klein/winding $(n_i;w_i)$ states    
in each of these sectors are those with $\sum_i n_iw_i=0,0,1,-1,2,$
and $-2$ respectively.  
Thus, proceeding sector by sector,
we find that $(M_\ast/M_{\rm string})^2=4/3$ in each sector. 
Thus, in each of these sectors, we conclude that {\it all}\/ of 
the Kaluza-Klein and winding states are heavier than the string scale.

Note that the level-matching constraint is critical in reaching this conclusion.
For example, the complete spectrum for a five-dimensional compactification 
on the $SO(10)$ torus in Eq.~(\ref{SO10torus}) is given by
\beq
     \alpha' M_{n_i;w_i}^2 ~=~ 0, ~2,~{5/2},~4,~{9/2},~6,~{13/2},~... 
\label{SO10spectrum}
\eeq
Clearly, our result would fail if a Kaluza-Klein/winding
state with $\alpha' M_{n_i;w_i}^2 = 2k+1/2$, $k\in\IZ$,
were to emerge in a sector for which 
$\alpha' M_{\rm tot}^2 = \alpha' M_{n_i;w_i}^2 -2$,
since this would yield a Kaluza-Klein/winding
state with $M_{\rm tot}^2/M_{\rm string}^2 = 1/2$.
Such a state would then be lighter than the string scale.
However, it is easy to show that the states with 
masses $\alpha' M_{n_i;w_i}^2 = 2k+1/2$ in Eq.~(\ref{SO10spectrum})
all have 
\beq
    \sum_i n_i w_i ~=~ \cases{ {\rm even} & if $k$ is odd \cr
                               {\rm odd} & if $k$ is even~. \cr}
\eeq
The level-matching constraint thus ensures that these states can only appear
in sectors for which $\alpha' M_{\rm tot}^2 = \alpha' M_{n_i;w_i}^2 +A$
where $A= -4, 0, 4, ...$.
Thus, once again, we see that all of the Kaluza-Klein states
and all of the winding states are heavier than the string scale.

Similar results also hold for the superstring.
In this case, there are four sectors:  the left- and right-moving
worldsheet fermions can be either Neveu-Schwarz (for which 
$a$ or $\overline{a} = -1/2$) or Ramond (for which   
$a$ or $\overline{a} = 0$).
Likewise, depending on this choice, the values of $N$ and $\Nbar$
can be restricted either to integers (Ramond) or to integers and half-integers 
(Neveu-Schwarz).
Working out the various combinations, we find that $\calM_{\rm tot}= \calM_{n_i;w_i}$
in all relevant sectors which are capable of yielding light states.  
The only exception is the tachyonic NS/NS sector
with $(N,\Nbar)=(0,0)$, but this sector is typically removed 
through the same GSO projection that introduces spacetime supersymmetry.
Thus, we again see that it is possible to choose compactification tori
such that all Kaluza-Klein states and all winding states 
are heavier than the string scale.  
Such results also generalize to the heterotic string
and (for Kaluza-Klein states) to 
the bulk sector of Type~I strings as well.

We conclude this section with several important comments and caveats.

First,  although our main result is that it is possible
to cross the string scale without seeing a single Kaluza-Klein or winding-mode
state,   we stress that this statement refers only to those
states which are {\it beyond}\/ the massless level.
Indeed, as we have seen, even at the massless level we generically
have states which are composed of non-trivial combinations of
Kaluza-Klein, winding, and oscillator excitations.
Such states may be even be part of the Standard Model 
gauge group and matter content.

Second, we have not yet discussed the scale for the oscillator
excitations.  In general, the mass scale for the oscillator   
modes is set by $M_{\rm string}$.  However, we see from Eq.~(\ref{constraints})
that each oscillator excitation contributes {\it two}\/ units to
the net value of $\alpha' M_{\rm tot}^2$.
Thus, even the string {\it oscillator}\/ modes
are generically heavier than the string scale!
Moreover, if we take the level-matching constraint into account,
the oscillator mass gap becomes even greater.  As an example,
let us consider the oscillator excitations of the graviton.  
In the bosonic string, 
the graviton emerges as a state with $(N,\Nbar)=(1,1)$ and $n_i=w_i=0$.
However, we see from the level-matching constraint in Eq.~(\ref{constraints})
that it is impossible to excite a single additional oscillator mode, since 
this would require changing the value of $\sum_i n_i w_i$ by introducing
simultaneous Kaluza-Klein/winding excitations. 
We therefore find that we 
need to excite oscillator modes in groups of two, keeping $N=\Nbar$.
Thus, the first oscillator excitation of the graviton 
does not appear until $\alpha' M_{\rm tot}^2 = 4$.

Third, although we have given examples where our results 
hold in all string sectors, it is undoubtedly possible to
construct string models where this fails to be the case.
In such instances, the nature of the lightest string states
may vary with the specific string sector in question.
For example, it may happen that the lightest compactification
states may exceed the string scale in one sector (\eg, for the
gauge bosons), yet be below this scale in another sector
(\eg, for quarks and leptons).  This would then give rise
to an interesting low-energy phenomenology in which different Standard
Model particles have Kaluza-Klein/winding excitations of different
masses, even though these excitations all correspond to the same extra 
dimensions with fixed radii!

Fourth, we emphasize that our discussion in this section has focused
only on those generic features which are common to all string theories.
As such, we have not focused on a particular class of string theories,
nor have we focused on the compactification of {\it all}\/ of the
extra spacetime dimensions predicted within a given class of string theories.  
In general, compactification of the full six (or seven) extra dimensions
predicted by string theory will give rise to additional constraints
beyond those which we have considered here.
(For example, modular invariance requires that the total theory
be compactified on tori for which the corresponding charge lattice
is even and self-dual~\cite{reviews,Narain}, so that 
$\alpha' M_{n_i;w_i}^2 \in 2\IZ$ only.)  
In this paper, by contrast, we are only focusing on a subset
of the full theory.  

Fifth, we remark that our results hold for all 
values of $M_{\rm string}$.  Thus, we have not focused on
the various mechanisms~\cite{TeV2,TeV3,TeV4,others}
by which the string scale might be
reduced into the TeV range in a variety 
of different classes of string theories.  
Of course, one possibility is that this reduction might occur as
a result of two or more extra dimensions taking values which are
significantly larger than the string length~\cite{TeV2}.   
Our results clearly do not apply for such extra dimensions.  
Instead, our focus here has been on those extra dimensions 
(so-called ``TeV-sized extra dimensions'')
whose sizes are relatively close to the fundamental string 
scale~\cite{TeV1,TeV3,TeV4,others,universal}. 
Indeed, it is only for these extra dimensions that our result
applies, and for which it is possible
to cross the string scale without seeing any corresponding
Kaluza-Klein and/or winding modes.

Finally, we note that in this paper we have been discussing
the properties of merely the tree-level string spectrum.
We have therefore been identifying the Standard Model
particles as massless string states, and focusing on light
excitations thereof.
Needless to say, field-theoretic effects such as electroweak symmetry
breaking and supersymmetry breaking, and even string-theoretic
effects such as vacuum restablization, have the potential
to shift the particle masses away from their tree-level 
values~\cite{phenoreviews}.
These effects can therefore be significant in cases where the   
string scale itself is in the TeV range.

\section{Conclusions}

In this paper, we have shown that there exist toroidal compactifications
for which the lightest Kaluza-Klein states {\it and}\/ the lightest
winding states are both simultaneously heavier than the string scale. 
The key ingredient in these compactifications is the presence of
non-trivial shape moduli and background antisymmetric tensor fields.

It is perhaps not surprising that non-trivial
shape moduli have the potential to alter some of our na\"\i ve
expectations concerning the masses of compactification states.
In Ref.~\cite{firstpaper}, for example, it was shown that such
moduli have the potential to render certain types of large extra
dimensions invisible.  Moreover, in Ref.~\cite{secondpaper},
it was shown that 
non-trivial shape moduli can trigger a so-called ``shadowing'' effect
in which compactification geometry, much like other ``constants'' of nature,
is effectively {\it renormalized}\/ as a function of energy scale,
with quantities such as compactification radii
changing their apparent values as functions of the energy
with which the compactification manifold is probed.
Results such as these suggest that shape moduli have the potential to
drastically change our na\"\i ve expectations based on studying
simple compactifications in which shape moduli are ignored or held fixed.

It is also important to stress that such compactifications are not
unusual in any way, and in fact are expected on rather general grounds.  
Even though we do not know the (presumably
non-perturbative) dynamics which ultimately selects the preferred 
compactification,
we expect that any effective potential which selects this preferred
compactification should reflect the underlying symmetries of the torus
and hence should be modular invariant.  It then follows that the extrema
of this potential should correspond to tori which sit at modular fixed points.  
However, tori such as the $SU(3)$ and $SO(10)$ tori discussed above 
correspond to highly symmetric modular fixed
points.  Indeed, as we have seen,
these are exactly the sorts of tori which give
rise to enhanced gauge symmetries in the full string theory.
It is therefore reasonable to assume that tori such 
as these are preferred dynamically.

Needless to say, these results could have important phenomenological implications 
in theories involving TeV-sized extra dimensions~\cite{TeV1,TeV3,TeV4,others,universal}. 
Ordinarily, one might have
assumed that the first experimental signals of such extra dimensions would be  
the appearance of Kaluza-Klein states, and that such Kaluza-Klein
states (or their winding counterparts) must necessarily appear 
below the string scale.
However, our results indicate that this need not be the case:
if certain extra dimensions are near the string scale,
it is possible to cross the string scale without seeing
a single corresponding state of either the Kaluza-Klein, winding,
or oscillator variety!
Thus, it is possible that the string scale may be lower than
previously imagined.

\section*{Acknowledgments}

This work was supported in part by the National Science Foundation
under Grant PHY-0071054, and by a Research Innovation Award from 
Research Corporation. 
 

\medskip



\begin{references}

\bibitem{TeV1}
       I.~Antoniadis, Phys.\ Lett.\ B {\bf 246} (1990) 377;
        I.~Antoniadis, K.~Benakli and M.~Quiros,
        Phys.\ Lett.\ B {\bf 331} (1994) 313
         [arXiv:hep-ph/9403290].

\bibitem{TeV2}
       N.~Arkani-Hamed, S.~Dimopoulos and G.~Dvali,
           Phys.\ Lett.\ B {\bf 429} (1998) 263
          [arXiv:hep-ph/9803315];
            I.~Antoniadis {\it et al.},
             Phys.\ Lett.\ B {\bf 436} (1998) 257
              [arXiv:hep-ph/9804398].

\bibitem{TeV3} K.~R.~Dienes, E.~Dudas and T.~Gherghetta,
           Phys.\ Lett.\ B {\bf 436} (1998) 55
          [arXiv:hep-ph/9803466];
           Nucl.\ Phys.\ B {\bf 537} (1999) 47
               [arXiv:hep-ph/9806292];
           arXiv:hep-ph/9807522.

\bibitem{TeV4} E.~Witten, Nucl.\ Phys.\ B {\bf 471} (1996) 135
             [arXiv:hep-th/9602070];
          J.D.~Lykken, Phys.\ Rev.\ D {\bf 54} (1996) 3693
            [arXiv:hep-th/9603133];
          G.~Shiu and S.-H.H.~Tye, Phys.\ Rev.\ D {\bf 58} (1998) 106007
             [arXiv:hep-th/9805157];
          Z.~Kakushadze and S.-H.H.~Tye,
              Nucl.\ Phys.\ B {\bf 548} (1999) 180
             [arXiv:hep-th/9809147].


\bibitem{missusy}  
        See, \eg, K.~R.~Dienes,
            Nucl.\ Phys.\ B {\bf 611} (2001) 146
           [arXiv:hep-ph/0104274];
              Nucl.\ Phys.\ B {\bf 429} (1994) 533
               [arXiv:hep-th/9402006];
         K.~R.~Dienes, M.~Moshe and R.~C.~Myers,
         Phys.\ Rev.\ Lett.\  {\bf 74} (1995) 4767
          [arXiv:hep-th/9503055].


\bibitem{Giveon}
        See, \eg, A.~Giveon, M.~Porrati and E.~Rabinovici,
               Phys.\ Rept.\  {\bf 244} (1994) 77
              [arXiv:hep-th/9401139].


\bibitem{nextpaper}   K.R.~Dienes and A.~Mafi, 
         {\it Phenomenological Implications of Compactifications on 
             Manifolds with Non-Trivial Shape Moduli}\/, to appear.


\bibitem{reviews}
       See, \eg, M.B. Green, J.H. Schwarz, and E. Witten, {\it Superstring Theory}\/
         (Cambridge University Press, 1987);
       J. Polchinski, {\it String Theory}\/ (Cambridge University Press, 1998).

\bibitem{phenoreviews}
      For phenomenological reviews, see, {\it e.g.}\/,
      L.E. Ib\'a\~nez, {\rm arXiv:hep-th/9112050}; {\rm arXiv:hep-th/9505098};
      I. Antoniadis, {\rm arXiv:hep-th/9307002};
      M. Dine, {\rm arXiv:hep-ph/9309319};
      A.E. Faraggi, {\rm arXiv:hep-ph/9405357};
      J.L. Lopez and D.V. Nanopoulos, {\rm arXiv:hep-ph/9511266};
      J.D. Lykken, {\rm arXiv:hep-ph/9511456};
      Z. Kakushadze and S.-H.H. Tye, {\rm arXiv:hep-th/9512155};
      F. Quevedo, {\rm arXiv:hep-th/9603074}; {\rm arXiv:hep-ph/9707434};\break
      K.R. Dienes, {\rm Phys.\ Rept.}\/ {\bf 287} (1997) 447;
      Z.~Kakushadze, G.~Shiu, S.-H.H.~Tye and Y.~Vtorov-Karevsky,
             Int.\ J.\ Mod.\ Phys.\ {\bf 13} (1998) 2551.




\bibitem{Narain} K.~S.~Narain, Phys.\ Lett.\ B {\bf 169} (1986) 41;
          K.~S.~Narain, M.~H.~Sarmadi and E.~Witten,
           Nucl.\ Phys.\ B {\bf 279} (1987) 369.


\bibitem{higherlevel} 
       K.~R.~Dienes and J.~March-Russell,
          Nucl.\ Phys.\ B {\bf 479} (1996) 113
          [arXiv:hep-th/9604112];
          K.~R.~Dienes, A.~E.~Faraggi and J.~March-Russell,
        Nucl.\ Phys.\ B {\bf 467} (1996) 44
        [arXiv:hep-th/9510223].


\bibitem{others}
     See, \eg,
   I.~Antoniadis and B.~Pioline, Nucl.\ Phys.\ B {\bf 550} (1999) 41
              [arXiv:hep-th/9902055];
   I.~Antoniadis, S.~Dimopoulos and A.~Giveon, JHEP {\bf 0105} (2001) 055
         [arXiv:hep-th/0103033].
 

\bibitem{universal}
    T.~Appelquist, H.~C.~Cheng and B.~A.~Dobrescu, 
     Phys.\ Rev.\ D {\bf 64} (2001) 035002 [arXiv:hep-ph/0012100].



\bibitem{firstpaper}   
         K.~R.~Dienes, Phys.\ Rev.\ Lett.\  {\bf 88} (2002) 011601
        [arXiv:hep-ph/0108115].

\bibitem{secondpaper}  
        K.~R.~Dienes and A.~Mafi, Phys.\ Rev.\ Lett.\  {\bf 88} (2002) 111602
        [arXiv:hep-th/0111264].






\end{references}
\end{document}